\documentclass[a4paper,11pt]{article}
\usepackage{pos}

\title{Weak-basis invariants and CP conservation in the leptonic sector with Majorana neutrinos}
\ShortTitle{Weak-basis invariants and CP conservation}

\author[a,b]{Bingrong Yu}
\author*[a,b]{Shun Zhou}

\affiliation[a]{Institute of High Energy Physics, Chinese Academy of Sciences,\\
  Beijing 100049, China}

\affiliation[b]{School of Physical Sciences, University of Chinese Academy of Sciences,\\
Beijing 100049, China}

\emailAdd{yubr@ihep.ac.cn}
\emailAdd{zhoush@ihep.ac.cn}

\abstract{In this talk, we present a recent investigation of the sufficient and necessary conditions for CP conservation in the leptonic sector with massive Majorana neutrinos in terms of CP-odd weak-basis invariants. The number of weak-basis invariants to guarantee CP conservation in the leptonic sector is clarified and a new set of invariants are advocated for the description of CP conservation, given the physical parameters in their experimentally allowed regions.}

\FullConference{
  40th International Conference on High Energy physics - ICHEP2020\\
  July 28 - August 6, 2020\\
  Prague, Czech Republic (virtual meeting)
}

\begin{document}
\maketitle

\section{Introduction}

The future long-baseline accelerator neutrino oscillation experiments are about to discover CP violation in the leptonic sector and precisely measure the relevant CP-violating phase~\cite{Branco:2011zb}. On the other hand, CP violation is an indispensable ingredient to dynamically generate cosmological matter-antimatter asymmetry~\cite{Sakharov:1967dj}. If leptonic CP violation is significantly large, then the matter-antimatter asymmetry in our Universe may be naturally explained via the leptogenesis mechanism~\cite{Fukugita:1986hr}. At the low-energy scale, the  effective Lagrangian with three massive Majorana neutrinos in the leptonic sector is given by
\begin{eqnarray}
{\cal L}^{}_{\rm lepton}=-\overline{l^{}_{\rm L}} M_{l}^{} l_{\rm R}^{} - \frac{1}{2}\overline{\nu^{}_{\rm L}}M_{\nu}^{} \nu_{\rm L}^{\rm C} + \frac{g}{\sqrt{2}}\overline{l_{\rm L}^{}}\gamma_{}^{\mu}\nu_{\rm L}^{}W_{\mu}^{-} +\rm h.c.\; ,
\label{eq:lag}
\end{eqnarray}
where $\nu_{\rm L}^{\rm C} \equiv {\cal C}\overline{\nu_{\rm L}^{}}_{}^{\rm T}$ with ${\cal C}\equiv {\rm i} \gamma^2 \gamma^0$ being the charge-conjugation matrix, while $M_{l}^{}$ and $M_{\nu}^{}$ denote the charged-lepton mass matrix and the Majorana neutrino mass matrix, respectively. The last term in Eq.~(\ref{eq:lag}) stands for the charged-current weak interaction with $g$ being the gauge coupling constant of the ${\rm SU}(2)_{\rm L}^{}$ gauge group. Leptonic flavor mixing and CP violation come from the mismatch between the diagonalization of the complex charged-lepton and Majorana neutrino mass matrices. When changing to the mass basis where $M_{l}^{} = \widehat{M}_{l}^{} \equiv {\rm Diag}\left\{m_e^{}, m_{\mu}^{}, m_{\tau}^{}\right\}$ and $M_{\nu}^{} = \widehat{M}_{\nu}^{} \equiv {\rm Diag}\left\{m_1^{}, m_2^{}, m_3^{}\right\}$, the leptonic flavor mixing matrix, also called the Pontecorvo-Maki-Nakagawa-Sakata (PMNS) matrix~\cite{Pontecorvo:1957cp, Maki:1962mu}, appears in the the charged-current interaction and leads to CP violation. In the standard parametrization~\cite{Zyla:2020zbs}, the PMNS matrix reads
\begin{eqnarray}
U^{}_{\rm PMNS}=\left( \begin{matrix} c^{}_{13} c^{}_{12} & c^{}_{13} s^{}_{12} & s^{}_{13} e^{-{\rm i}\delta} \cr -s_{12}^{} c_{23}^{} - c_{12}^{} s_{13}^{} s_{23}^{} e^{{\rm i}\delta}_{} & + c_{12}^{} c_{23}^{} - s_{12}^{} s_{13}^{} s_{23}^{} e^{{\rm i}\delta}_{} & c_{13}^{} s_{23}^{} \cr + s_{12}^{} s_{23}^{} - c_{12}^{} s_{13}^{} c_{23}^{} e^{{\rm i}\delta}_{} & - c_{12}^{} s_{23}^{} - s_{12}^{} s_{13}^{} c_{23}^{} e^{{\rm i}\delta}_{} & c_{13}^{} c_{23}^{} \end{matrix} \right) \cdot \left(\begin{matrix} e^{{\rm i}\rho} & 0 & 0 \cr 0 & e^{{\rm i}\sigma} & 0 \cr 0 & 0 & 1\end{matrix}\right) \; ,
\label{eq:PMNS}
\end{eqnarray}
where $c^{}_{ij} \equiv \cos \theta^{}_{ij}$ and $s^{}_{ij} \equiv \sin \theta^{}_{ij}$ (for $ij = 12, 13, 23$) have been defined, $\{\theta^{}_{12}, \theta^{}_{13}, \theta^{}_{23}\}$ are three flavor mixing angles, and $\{\delta, \rho, \sigma\}$ are three CP-violating phases.

In order to describe CP violation and find out the sufficient and necessary conditions for CP conservation in a basis-independent way, one can construct the CP-odd weak-basis (WB) invarants following the idea first put forward by Branco and his collaborators~\cite{Bernabeu:1986fc, Branco:1986gr}. In the leptonic sector, CP is conserved if and only if the Lagrangian in Eq.~(\ref{eq:lag}) is invariant under the following generalized CP transformation \cite{Branco:1986gr}
\begin{eqnarray}
l^{}_{\rm L} \to  U^{}_{\rm L} {\cal C}l^*_{\rm L} \; , \quad \nu^{}_{\rm L} \to U^{}_{\rm L} {\cal C}\nu^*_{\rm L} \; , \quad
l^{}_{\rm R} \to  U^{}_{\rm R} {\cal C}l^*_{\rm R} \; , \quad
W^-_\mu \to  -(-1)^{\delta^{}_{0\mu}} W^+_\mu \; ,
\label{eq:GCP transformation}
\end{eqnarray}
where the asterisk ``$*$'' indicates the complex conjugation and $\delta^{}_{0\mu}$ (for $\mu = 0, 1, 2, 3$) stands for the Kronecker delta, while $U^{}_{\rm L}$ and $U^{}_{\rm R}$ are two arbitrary $3\times 3$ unitary matrices in the flavor space. Then the sufficient and necessary conditions for CP conservation in the leptonic sector are equivalent to the existence of the unitary matrices $U^{}_{\rm L}$ and $U^{}_{\rm R}$ satisfying~\cite{Branco:1986gr}
\begin{eqnarray}
U^\dagger_{\rm L} M^{}_\nu U^*_{\rm L} = - M^*_\nu \; , \quad U^\dagger_{\rm L} M^{}_l U^{}_{\rm R} = M^*_l \;.
\label{eq:mass matrix transformation}
\end{eqnarray}
In terms of $H_{l}^{}\equiv M_{l}^{}M_{l}^{\dagger}$, $H_{\nu}^{}\equiv M_{\nu}^{}M_{\nu}^{\dagger}$ and $G_{l\nu}^{}\equiv M_{\nu}^{}H_{l}^{*}M_{\nu}^{\dagger}$, it has been suggested in Ref.~\cite{Dreiner:2007yz} that the minimal set of sufficient and necessary conditions for CP conservation in the leptonic sector are given by the following three identities
\begin{eqnarray}
{\cal I}^{}_1 &\equiv& {\rm Tr}\left\{ \left[H^{}_\nu, H^{}_l \right]^3\right\} = 0 \; ,
 \label{eq:I1} \\
{\cal I}^{}_2 &\equiv& {\rm Im}\left\{{\rm Tr}\left[H^{}_l H^{}_\nu G^{}_{l\nu}\right]\right\} = 0 \; ,\label{eq:I2} \\
{\cal I}^{}_3 &\equiv& {\rm Tr}\left\{ \left[G^{}_{l\nu}, H^{}_l \right]^3\right\} = 0 \; . \label{eq:I3}
\end{eqnarray}
Note that all the masses of charged leptons and neutrinos are assumed to be non-degenerate, which is consistent with experimental observations~\cite{Zyla:2020zbs}. With the help of the transformations of $M_{l}^{}$ and $M_{\nu}^{}$ in Eq.~(\ref{eq:mass matrix transformation}), one can obtain $U^\dagger_{\rm L} H^{}_l U^{}_{\rm L} = H^*_l$, $U^\dagger_{\rm L} H^{}_\nu U^{}_{\rm L} = H^*_\nu$ and $U^\dagger_{\rm L} G^{}_{l\nu} U^{}_{\rm L} = G^*_{l\nu}$, and verify that ${\cal I}^{}_i$ (for $i = 1, 2, 3$) are indeed WB invariants and they all vanish if CP is conserved. On the other hand, it has been proved in Ref.~\cite{Branco:1986gr} that the vanishing of another set of four WB invariants serves as sufficient and necessary conditions for CP conservation with non-degenerate lepton masses, i.e.,
\begin{eqnarray}
\widehat{\cal I}^{}_1 &\equiv& {\rm Im}\left\{{\rm Tr}\left[H^{}_l H^{}_\nu G^{}_{l\nu}\right]\right\} = 0 \; ,
 \label{eq:Ip1} \\
\widehat{\cal I}^{}_2 &\equiv& {\rm Im}\left\{{\rm Tr}\left[H^{}_l H^2_\nu G^{}_{l\nu}\right]\right\} = 0 \; ,\label{eq:Ip2} \\
\widehat{\cal I}^{}_3 &\equiv& {\rm Im}\left\{{\rm Tr}\left[H^{}_l H^2_\nu G^{}_{l\nu} H^{}_\nu \right]\right\} = 0 \; , \label{eq:Ip3} \\
\widehat{\cal I}^{}_4 &\equiv& {\rm Im}\left\{{\rm Det}\left[G^{}_{l\nu} + H^{}_l H^{}_\nu \right]\right\} = 0 \; . \label{eq:Ip4}
\end{eqnarray} 
An immediate question is whether the number of sufficient and necessary conditions for leptonic CP conservation is three or four.

In this talk, motivated by this question, we reexamine the sufficient and necessary conditions for leptonic CP conservation with massive Majorana neutrinos. An intuitive answer would be three, which is also the total number of independent phases in the PMNS matrix in Eq.~(\ref{eq:PMNS}). However, as we shall show by a counter example, the vanishing of all the three invariants ${\cal I}_{i}$ (for $i=1, 2, 3$) is in general \emph{not} sufficient to guarantee CP conservation. Moreover, we suggest a new set of three WB invariants, and demonstrate that the vanishing of them indeed serves as the sufficient and necessary conditions for CP conservation at least in the whole experimentally allowed parameter space of lepton masses and mixing angles~\cite{Yu:2019ihs}.

\section{Sufficient and necessary conditions}

Now we show a counter example for the statement that ${\cal I}^{}_i = 0$ (for $i = 1, 2, 3$) are sufficient for CP conservation~\cite{Yu:2019ihs}. First, in the basis where both $M_l^{}$ and $M_{\nu}^{}$ are diagonal, one can explicitly calculate ${\cal I}^{}_1$ by using the PMNS matrix in Eq.~(\ref{eq:PMNS}). It turns out that
\begin{eqnarray}
{\cal I}_1^{} = - 6 {\rm i} \Delta^{}_{21} \Delta^{}_{31} \Delta^{}_{32} \Delta^{}_{e\mu} \Delta^{}_{\mu\tau} \Delta^{}_{\tau e} {\cal J} \; , \label{eq:I1expr}
\end{eqnarray}
where $\Delta^{}_{ij} \equiv m^2_i - m^2_j$ (for $i, j = 1, 2, 3$) and $\Delta^{}_{\alpha \beta} \equiv m^2_\alpha - m^2_\beta$ (for $\alpha, \beta = e, \mu, \tau$) denote the mass-squared differences for neutrinos and charged leptons, respectively. In addition, the Jarlskog invariant~\cite{Jarlskog:1985ht, Wu:1985ea} is given by ${\cal J} \equiv s^{}_{12} c^{}_{12} s^{}_{23} c^{}_{23} s^{}_{13} c^2_{13} \sin\delta $. Given the best-fit values of neutrino mass-squared differences $\{\Delta^{}_{21} = 7.39\times 10^{-5}~{\rm eV}^2, \Delta^{}_{31} = 2.523\times 10^{-3}~{\rm eV}^2\}$ and three mixing angles $\{\theta^{}_{12} = 33.82^\circ, \theta^{}_{13} = 8.61^\circ, \theta^{}_{23} = 48.3^\circ\}$ from Ref.~\cite{Esteban:2018azc},\footnote{Note that we take normal neutrino mass ordering for illustration, and the case of inverted mass ordering can be analyzed in a similar way.} together with the charged-lepton masses $m^{}_e = 0.511~{\rm MeV}$, $m^{}_{\mu} = 105.658~{\rm MeV}$ and $m^{}_{\tau} = 1776.86~{\rm MeV}$~\cite{Zyla:2020zbs}, one can see that ${\cal I}_{1}^{}$ vanishes if and only if the Dirac CP phase $\delta$ takes a trivial value (i.e., $0$ or $\pi$). As long as ${\cal I}_{1}^{}=0$ holds, ${\cal I}^{}_{2} = 0$ and ${\cal I}^{}_{3} = 0$ give two nonlinear equations of the Majorana CP phases $\rho$ and $\sigma$, namely,
\begin{eqnarray}
{\cal I}_2^{\prime} &=& -2.092 \sin(2\rho) -8.754 \sin(2\sigma) - 0.035 \sin(2\rho-2\sigma) = 0 \; , \label{eq:I2num}\\
{\cal I}_3^{\prime} &=& +0.471 \sin(2\rho) - 3.535 \sin(2\sigma) + 1.177 \sin(2\rho-2\sigma) \nonumber \\
&~&  - 0.199 \sin(2\rho+2\sigma) - 2.574 \sin(2\rho-4\sigma) - 0.934 \sin(2\sigma-4\rho) = 0 \; , \label{eq:I3num}
\end{eqnarray}
where the invariants ${\cal I}_2^{\prime} \equiv {\cal I}^{}_2 /(10^6~{\rm eV}^4 \cdot {\rm MeV}^4)$ and ${\cal I}_3^{\prime} \equiv {\cal I}^{}_3 /(6{\rm i}\cdot 10^{24}~{\rm eV}^6 \cdot {\rm MeV}^{12})$ have been properly normalized. In Eqs.~(\ref{eq:I2num}) and ~(\ref{eq:I3num}), the best-fit values of all the physical parameters have been taken, except for CP phases. In particular, we have assumed the yet-unknown lightest neutrino mass to be $m^{}_1 = 0.03~{\rm eV}$ just for illustration, which is well compatible with the cosmological bound on neutrino masses~\cite{Aghanim:2018eyx}.\footnote{The precision observations of the cosmic microwave background and the large-scale structures in our Universe give the upper bound $m^{}_1 + m^{}_2 + m^{}_3 < 0.12~{\rm eV}$ on neutrino masses, implying $0 \leq m^{}_1 < 0.04~{\rm eV}$ in the case of normal neutrino mass ordering.} Apart from the trivial solutions (i.e., $0$ or $\pi/2$), Eqs.~(\ref{eq:I2num}) and ~(\ref{eq:I3num}) $\emph{do}$ have nontrivial solutions for $\rho$ and $\sigma$, indicating the existence of CP violation,
\begin{eqnarray} 
\left\{
\begin{aligned}
\rho&=38.551^{\circ}\\
\sigma&=173.146^{\circ}\\
\end{aligned}
\right. \; ,
\qquad \text{or}\qquad
\left\{
\begin{aligned}
\rho&=141.449^{\circ}\\
\sigma&=6.854^{\circ}\\
\end{aligned}
\right. \; .
\label{eq:nontrivial sol}
\end{eqnarray}
This numerical counter example proves that ${\cal I}_i = 0$ (for $i=1, 2, 3$) are not sufficient conditions for leptonic CP conservation.

Generally speaking, since the vanishing of three WB invariants supplies three independent but nonlinear equations of three CP phases, the trivial values of three CP phases are not the unique solution to these equations. Certainly the nontrivial solutions depend on all the physical parameters. Thanks to the elegant neutrino oscillation experiments, we now have quite a precise knowledge about neutrino mass-squared differences and flavor mixing angles. Furthermore, the charged-lepton masses are precisely measured, so the only unknown parameter that the nontrivial solutions of CP phases depend on is the absolute mass of the lightest neutrino $m^{}_1$. This allows us to find another set of three WB invariants such that the vanishing of these invariants renders all the CP phases to take only trivial values, at least in the physical allowed parameter space, i.e., $0 \leq m^{}_1 < 0.04 ~{\rm eV}$. The construction of new WB invariants is straightforward, given the transformation rules of $H_{l}^{}$, $H_{\nu}^{}$ and $G_{l\nu}^{}$ under the WB transformation. One can write down a series of WB invariants in the form of ${\cal I}^{kmn}_{rst} \equiv {\rm Im}\left\{{\rm Tr}\left[H^{k}_l H^{m}_\nu G^{n}_{l\nu} H^{r}_l H^{s}_\nu G^{t}_{l\nu}\cdots\right]\right\}$, where the power indices $\{k, m, n, r, s, t\}$ are non-negative integers and the ellipsis denotes the additional matrices composed of $H^{}_l$, $H^{}_\nu$ and $G^{}_{l\nu}$. It is easy to verify these quantities ${\cal I}^{kmn}_{rst}$ are invariant under the WB transformations and they are related to all three CP-violating phases.

\begin{figure}[!t]
\centering	
\includegraphics[width=0.49\textwidth]{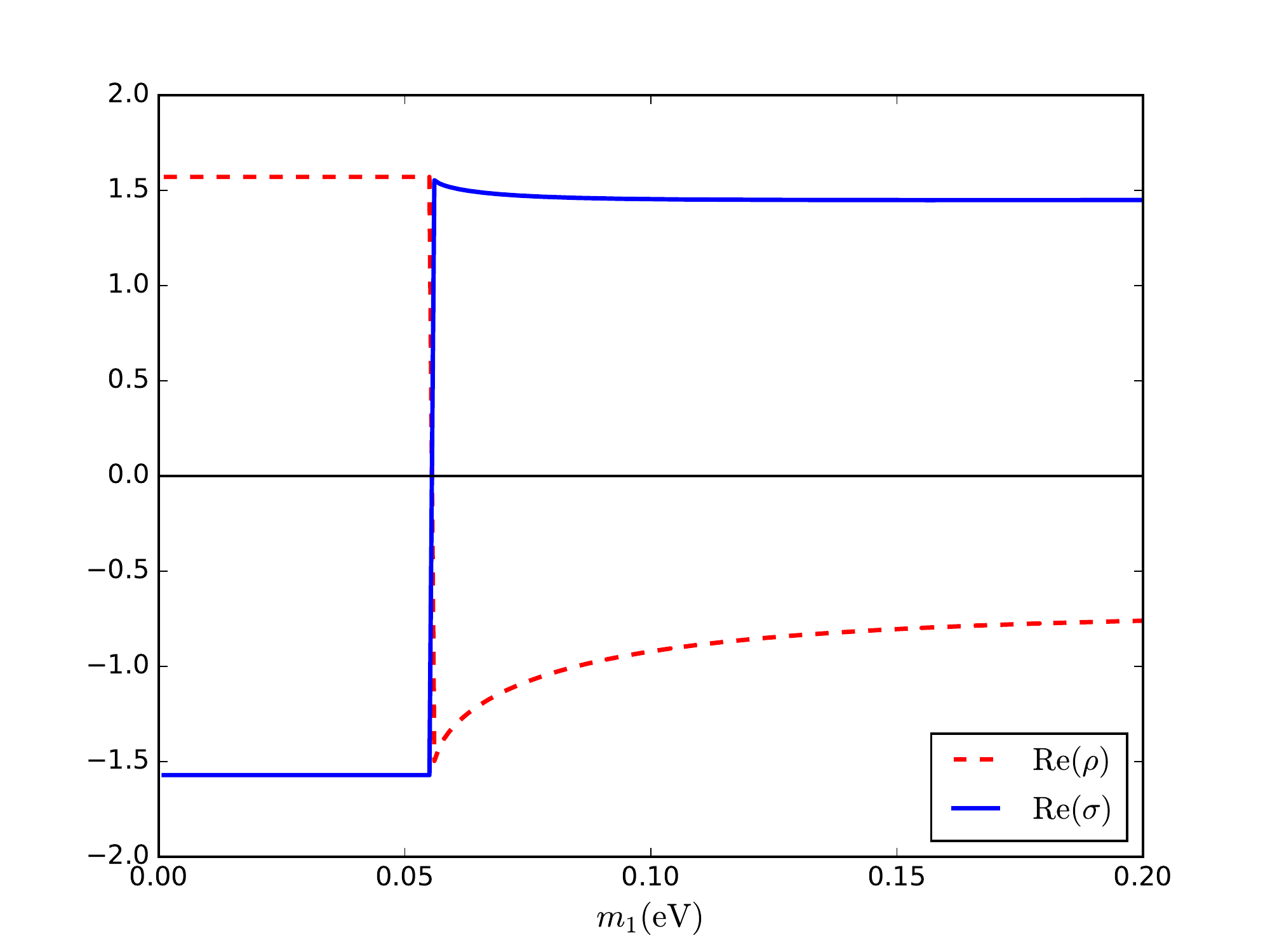}	
\includegraphics[width=0.49\textwidth]{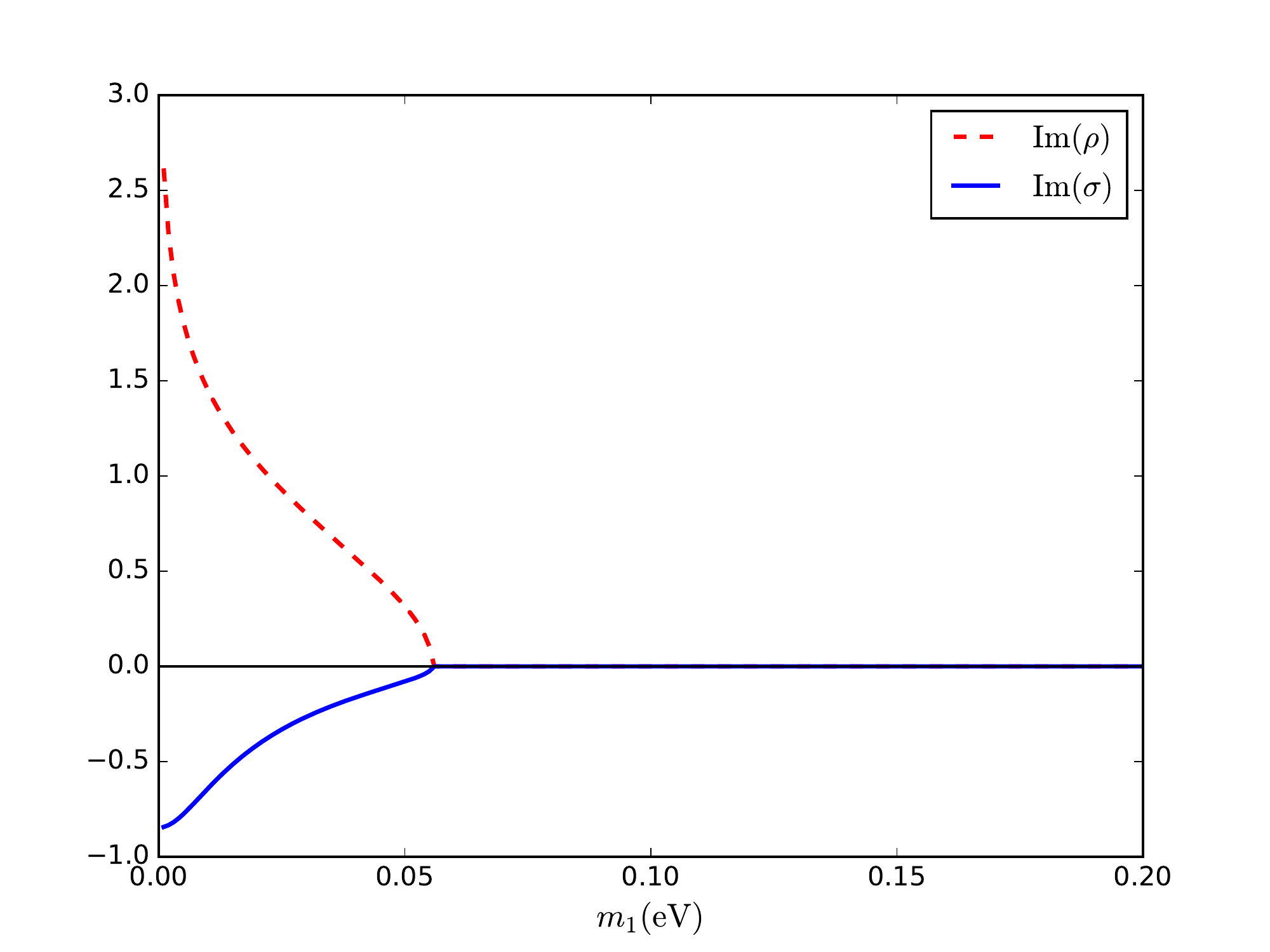}	
\caption{Given ${\cal I}_1^{} = {\cal I}_2^{} = \widehat{\cal I}_2^{} = 0$, the real parts (left panel) and imaginary parts (right panel) of the nontrivial solutions of $\rho$ (red dashed curve) and $\sigma$ (blue solid curve) are shown as functions of $m^{}_1$.}
\label{fig:realimag}
\end{figure}

We find that it is indeed possible to construct a new set of three WB invariants, the vanishing of which guarantees CP conservation in the the experimentally allowed parameter space. For illustration, we choose $\{{\cal I}^{}_1, {\cal I}^{}_2, \widehat{\cal I}^{}_2\}$. As before, ${\cal I}^{}_1=0$ is equivalent to $\delta=0$ or $\pi$. After $\delta$ being set to trivial values, ${\cal I}^{}_2 = 0$ and  $\widehat{\cal I}^{}_2 = 0$ supply two independent and nonlinear equations of $\rho$ and $\sigma$. Then, we fix all the other physical parameters at their best-fit values mentioned before. Finally, for a given value of $m^{}_1$, we can numerically solve ${\cal I}^{}_2=\widehat{\cal I}^{}_2=0$. In Fig.~\ref{fig:realimag}, we show the real and imaginary parts of the nontrivial solutions of $\rho$ and $\sigma$ as functions of $m_1^{}$ in the left and right panel, respectively. Interestingly, both panels of Fig.~\ref{fig:realimag} give a critical point located at
\begin{eqnarray}
m_1^{} = m^{}_{*} \approx 0.0557~{\rm eV} \; ,
\label{eq:critical point}
\end{eqnarray}
whose exact value certainly depends on the input values of other physical parameters. For $m^{}_1 > m^{}_*$, the nontrivial solutions of $\rho$ and $\sigma$ are real and thus physically allowed, so the CP conservation is violated. For $m^{}_1 \leq m^{}_*$, the solutions of $\rho$ and $\sigma$ become purely imaginary, which is physically meaningless, so the CP conservation is maintained. As the cosmological bound on neutrino masses enforces $m_1^{}$ to be less than $0.04~{\rm eV}$, it is convincing to assert that  ${\cal I}^{}_1={\cal I}^{}_2= \widehat{\cal I}^{}_2=0$ is sufficient to guarantee CP conservation for all the physical parameters within their experimentally allowed regions. It is worthwhile to notice that the choice of three WB invariants $\{{\cal I}_1^{}, {\cal I}_2^{}, \widehat{\cal I}_2^{}\}$ is by no means unique, but the impact of different choices is just a different value of the critical mass $m_*^{}$. All we need to care about is to find a set of three WB invariants with the critical mass $m^{}_*$ greater than $0.04 ~\rm eV$. For example, the set $\{{\cal I}_1^{}, \widehat{\cal I}_2^{}, \widehat{\cal I}_3^{}\}$ also works, with the critical mass $m^{}_* \approx 0.142~{\rm eV}$.

\section{Summary}

We have found a counter example, in which three WB invariants given in Ref.~\cite{Dreiner:2007yz} vanish but CP violation is still present. Moreover, we advocate a new set of three WB invariants and prove that the vanishing of them serves as the sufficient and necessary conditions for CP conservation in the whole physically allowed parameter space. When neutrino masses are partially or completely degenerate, the number of CP phases and the sufficient and necessary conditions for CP conservation in the leptonic sector have recently been studied in Ref.~\cite{Yu:2020gre}.

\section*{Acknowledgements}

The authors thank Prof. Thomas Schwetz and Prof. Zhi-zhong Xing for helpful discussions. This work was supported in part by the National Natural Science Foundation of China under grant No.~11775232 and No.~11835013, and by the CAS Center for Excellence in Particle Physics.

\end{document}